\documentclass[conference]{IEEEtran}

\usepackage{cite}
\usepackage{amsmath,amssymb,amsfonts}
\usepackage{algorithmic}
\usepackage{graphicx}
\usepackage{textcomp}
\usepackage{xcolor}
\usepackage{enumitem}
\usepackage{outlines}
\usepackage{tabularx}
\usepackage{booktabs}
\usepackage{url}
\usepackage{textgreek}
\usepackage[ruled,vlined,linesnumbered]{algorithm2e}
\usepackage{bm}
\usepackage{float}
\usepackage{caption}
\usepackage{subcaption}

\DeclareMathOperator*{\argmax}{argmax}

\SetCommentSty{commentstyle}

\begin{document}

\title{Massively Parallel Causal Inference of Whole Brain Dynamics at Single Neuron Resolution}

\makeatletter
\newcommand{\linebreakand}{%
  \end{@IEEEauthorhalign}
  \hfill\mbox{}\par
  \mbox{}\hfill\begin{@IEEEauthorhalign}
}
\makeatother

\author{
    \IEEEauthorblockN{
        Wassapon Watanakeesuntorn\IEEEauthorrefmark{1}, 
        Keichi Takahashi\IEEEauthorrefmark{1}\textsuperscript{\textsection},
        Kohei Ichikawa\IEEEauthorrefmark{1}, 
        Joseph Park\IEEEauthorrefmark{2}, \\
        George Sugihara\IEEEauthorrefmark{3}, 
        Ryousei Takano\IEEEauthorrefmark{4}, 
        Jason Haga\IEEEauthorrefmark{4}, 
        Gerald M. Pao\IEEEauthorrefmark{5}\textsuperscript{\textsection}
    }
    \\
    \IEEEauthorblockA{\IEEEauthorrefmark{1} Nara Institute of Science and Technology, Nara, Japan, \\\{wassapon.watanakeesuntorn.wq0, keichi, ichikawa\}@is.naist.jp}
    \IEEEauthorblockA{\IEEEauthorrefmark{2} U.S. Department of the Interior, Florida, USA, \\josephpark@ieee.org}
    \IEEEauthorblockA{\IEEEauthorrefmark{3} University of California San Diego, California, USA, \\gsugihara@ucsd.edu}
    \IEEEauthorblockA{\IEEEauthorrefmark{4} National Institute of Advanced Industrial Science and Technology, Tsukuba, Japan, \\\{takano-ryousei, jh.haga\}@aist.go.jp}
    \IEEEauthorblockA{\IEEEauthorrefmark{5} Salk Institute for Biological Studies, California, USA, \\pao@salk.edu}
}

\IEEEoverridecommandlockouts
\IEEEpubid{\makebox[\columnwidth]{978-1-7281-9074-7/20/\$31.00~\copyright2020 IEEE \hfill} \hspace{\columnsep}\makebox[\columnwidth]{ }}

\maketitle

\begingroup\renewcommand\thefootnote{\textsection}
\footnotetext{Corresponding author}
\endgroup

\IEEEpubidadjcol

\begin{abstract}

Empirical Dynamic Modeling (EDM) is a nonlinear time series causal inference framework. The latest implementation of EDM, cppEDM, has only been used for small datasets due to computational cost. With the growth of data collection capabilities, there is a great need to identify causal relationships in large datasets. We present mpEDM, a parallel distributed implementation of EDM optimized for modern GPU-centric supercomputers. We improve the original algorithm to reduce redundant computation and optimize the implementation to fully utilize hardware resources such as GPUs and SIMD units. As a use case, we run mpEDM on AI Bridging Cloud Infrastructure (ABCI) using datasets of an entire animal brain sampled at single neuron resolution to identify dynamical causation patterns across the brain. mpEDM is 1,530$\times$ faster than cppEDM and a dataset containing 101,729 neuron was analyzed in 199 seconds on 512 nodes. This is the largest EDM causal inference achieved to date.

\end{abstract}


\section{Introduction }\label{introduction}


Reverse-engineering and building a digital reconstruction of the brain is one 
of the greatest scientific challenges of today. A recent study on the mouse 
cortex \cite{guamuanuct2018mouse} showed that 97\% of the possible connections 
between neurons exist. This result suggests that it is likely more informative to
investigate the dynamic interactions between neurons rather than the static
connectivity between them to fully understand the function of the brain. Based
on this insight, we are building mathematical and computational tools to
analyze the dynamic interactions between neurons based on Empirical Dynamic
Modeling (EDM). 


EDM is a nonlinear time series causal inference framework
based on the generalized Takens' embedding theorem on state space
reconstruction~\cite{deyle2011generalized}. EDM is used to study and predict the
behavior of nonlinear dynamical systems. Convergent Cross Mapping (CCM) is one of the EDM
algorithms that allows to estimate the existence and strength of the causal strength between two
time series in a dynamical system~\cite{sugihara2012detecting}. 

In this study, we utilize CCM to infer the causal
relationships between every neuron in an entire brain and
construct a causal map that describes the dynamic interactions among neurons.
For this purpose, we have recorded the neural activity (\emph{i.e.} firing rate) of an
entire larval zebrafish brain at singe-neuron resolution by using light sheet
fluorescence microscopy.
The original implementation of EDM, cppEDM, has mostly been used for individual time series of
relatively short length and and mostly small numbers of variables for its computational cost.
Since a larval zebrafish brain contains approximately $10^5$ neurons, a
staggering number of $10^{10}$ cross mappings need to be performed in total.
CCM of this enormous scale has never been achieved so far because of
the sheer amount of computation required.



The goal of this paper is to develop a highly scalable and optimized
implementation of EDM that is able to analyze the whole
zebrafish brain dataset within a reasonable time. 
We present mpEDM\footnote{\url{https://github.com/keichi/mpEDM}}, a parallel distributed implementation of EDM optimized for execution on modern GPU-centric supercomputers.
We improve the original algorithm in cppEDM to reduce redundant computation and optimize the implementation to fully utilize hardware 
resources such as GPUs and SIMD units.


Our evaluation on AI Bridging Cloud Infrastructure (ABCI), Japan's most high performance supercomputer as of today, 
demonstrated the unprecedented performance of mpEDM. mpEDM
was used to analyze a dataset containing the activity of 53,053 neurons in only 20
seconds using 512 ABCI nodes. In contrast, cppEDM took 8.5 hours to analyze
the same dataset using the same number of nodes~\cite{park2019massively}. Furthermore, mpEDM analyzed a
larger dataset containing 101,729 neurons in 199 seconds on 512 nodes. To our
knowledge, this is the largest CCM calculation achieved to date. This result
shows the potential for mpEDM and ABCI to analyze even larger datasets in the
future.


The rest of this paper is structured as follows. Section~\ref{background}
describes the background of this research and EDM algorithm.
Section~\ref{proposal} explains our proposal to improve the algorithm of the
EDM for parallelization and to support GPU architecture.
Section~\ref{evaluation} evaluates the performance of mpEDM and presents the
scientific outcomes obtained with mpEDM. Finally, section~\ref{summary}
concludes this paper and discusses future work.

\section{Background}\label{background}

\subsection{Causal Map of the Zebrafish Brain at Single Neuron Resolution}


To understand the human brain activity dynamics with a complexity of $10^{11}$ neurons and 
$10^{15}$ synapses at single neuron resolution is currently a technically impossible task. 
Similarly a mouse brain with $7.6\times10^{7}$ neurons is not tractable 
because mammalian brains are opaque and it is impossible to image a complete mouse brain. 
With this in mind, the zebrafish embryo is an attractive model system with 120,000 
neurons and transgenic technology as well as natural brain transparency. The zebrafish embryo 
is sufficiently complex to exhibit interesting behaviors and is technologically feasible to 
study to infer basic principles of systems neuroscience. Even in the case of the larval 
zebrafish with about 120,000 neurons we do not have the physical connectivity map, that is 
the connectome of the larval zebrafish, nor do we have the synaptic strengths which are pieces 
of information required to understand the brain starting from the physical connectivity.

Complicating this notion, recent work from the mouse brain shows that 97\% of possible 
physical connections exist within the mouse cortex thus making it difficult to analyze. 
Given this difficulty, using an analogy of a city; to understand how a city works it will 
be easier to understand the city from the traffic patterns than from the street map. Thus, 
we wished to analyze the fish brain at single neuron resolution from a network activity 
dynamics perspective. Although imperfect, we used neural activity imaging data of an entire 
brain at single cell resolution in a behaving larval zebrafish (a transparent vertebrate) 
to extract all relationships in an intact vertebrate brain. 

To achieve this, we recorded whole brain neural activity 
patterns in multiple animals experiencing hypoxia using a Selective Plane Illumination Microscope (SPIM)~\cite{ahrens2013whole}.  
We obtained data from the entire 5-day-old larval brain (120,000 neurons) at 2 Hz in response to hypoxia for varying amounts 
of time typically ranging from 1,500 time steps to up to 8,000+~\cite{chen2018brain}. 

CCM allows the inference of causation from nonlinear time series even with substantial noise and complete absence 
of correlation~\cite{clark2015spatial, ye2015distinguishing}. We used CCM and other tools from the EDM framework 
for the inference of existence, strength and sign of causal relationships within the neural activity network 
of the transparent larval fish brain~\cite{ahrens2013whole}. CCM determines whether and how much causality exists 
between individual neurons. The adjacency in the network is determined by time delay cross mapping~\cite{ye2015distinguishing}.
Predictive accuracy values give the interaction strength allow us to infer relationships within the 
neural network without observing the physical connectivity. As a test case, we have collected 
multiple data sets of lengths around 1600 time steps at 2 Hz which contain 50,000--80,000 active 
neurons in most cases. We have analyzed this data and show that the generated time series are 
suitable for causal network inference using the EDM framework and thus demonstrated a proof of 
principle of computational tractability.

\subsection{Empirical Dynamic Modeling} \label{Empirical Dynamic Modeling}

\begin{figure}
    \centering
    \centerline{\includegraphics[width=1\linewidth]{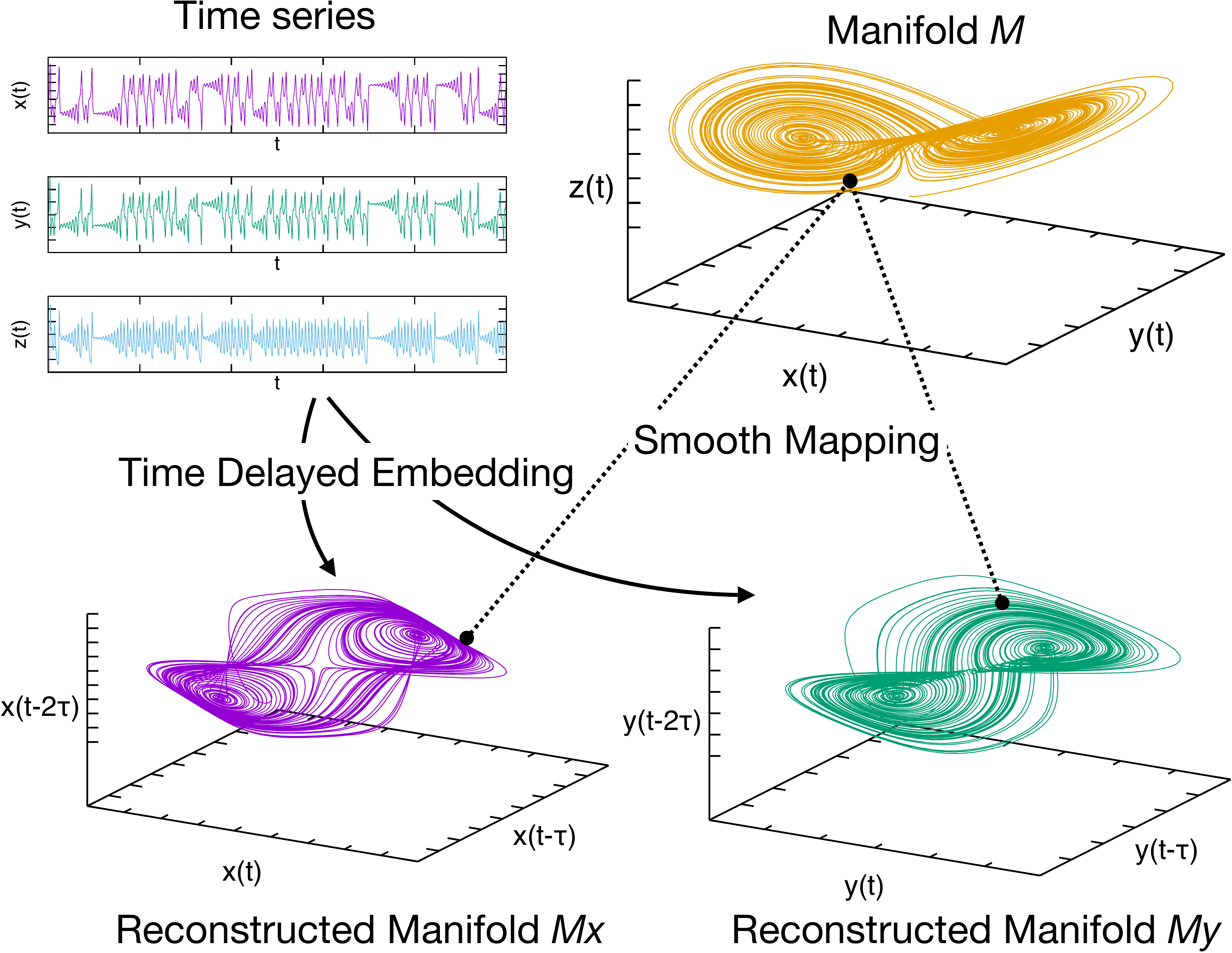}}
    \caption{Basic Idea Behind Empirical Dynamic Modeling}
    \label{fig:cross_mapping}
\end{figure}
    
EDM is a mathematical framework designed for
studying nonlinear dynamical systems. EDM is based upon the concept of state
space reconstruction (SSR)~\cite{sugihara1990nonlinear}. Takens' theorem states
that the attractor manifold of a multivariate dynamical system can be reconstructed from
time lagged coordinates of a single time series variable~\cite{takens1981detecting}.
Figure~\ref{fig:cross_mapping} illustrates the concept of state spaces
reconstruction. In this example, three causally related time series variables
$x(t)$, $y(t)$ and $z(t)$ that constitute a dynamical system form an attractor
manifold $M$ in the state space. A shadow manifold $M_x$ can be reconstructed
using the time delayed embeddings of $x$ ($x(t), x(t-\tau), x(t-2\tau),
\dots$), where $\tau$ denotes the time lag. In the same manner, lags of $y$ form a shadow manifold $M_y$. Takens'
theorem states that the reconstructed manifolds $M_x$ and $M_y$ preserve
essential mathematical properties (such as the topology) of the true manifold
$M$. In particular, there exist \textit{smooth} mappings between $M$, $M_x$,
and $M_y$, suggesting that neighbors in $M_x$ are neighbors in $M_y$ as
well.

Simplex projection is a nonlinear forecasting algorithm often used for
estimating the dimensionality of a dynamical system. In simplex projection,
the input time series is split into two halves: library $x$ and target $y$.
Both halves are embedded into $E$-dimensional state space by using delayed
embeddings. Given a point $\bm{y}(t_p) = (y(t_p), y(t_p-1), \dots,
y(t_p-E+1))$ in the target state space, its $E+1$ nearest neighbors (\emph{i.e.}
vertices of the simplex enclosing $\bm{y}(t_p)$) are searched from the embedded
library. Suppose those neighbors are $\bm{x_1}(t_1), \bm{x_2}(t_2), \dots,
\bm{x}_{E+1}(t_{E+1})$. A forecast $\bm{y}(t_p+1)$ can be made by averaging
the future of the neighbors in the library: $\bm{x_1}(t_1+1), \bm{x_2}(t_2+1),
\dots, \bm{x}_{E+1}(t_{E+1}+1)$. This prediction is performed for every point
in $\bm{y}$ and the results are compared with the true $y$ to evaluate the
prediction accuracy. This entire procedure is repeated for different $E$
values and the $E$ that achieves the highest prediction accuracy is determined
as the optimal embedding dimension of the dynamical system.

CCM determines the existence and strength of
causality between two time series variables~\cite{schiecke2015convergent}. It
works similar to simplex projection, but instead of predicting within a single
time series, CCM predicts one time series from another. If $y$ can be predicted
from $x$ with significant accuracy, we conclude that $y$ \textit{CCM causes} $x$.

There have been extensive studies on causal inference. Structural Causal Model (SCM) is one of the most popular causal models~\cite{pearl2009causal} based on statistical modeling of equilibrium systems. In contrast to SCM, EDM is based on the principle of state-space reconstruction shown in Takens’ theorem of non-equilibrium systems. Granger causality is another causal inference technique based on statistical modeling~\cite{granger1969investigating}. Granger causality however as stated by Granger himself, only works with linear and stochastic systems and cannot be applied to a nonlinear dynamical system. Compared to these alternatives, EDM is better suited to find the causal relationships in a nonlinear dynamical system such as the brain. Tajima \emph{et al.}~\cite{tajima2015untangling} also applied embedding theorems in nonlinear state-space reconstruction to analyze a dynamic system. They also built on the causality inference method from Sugihara \emph{et al.}~\cite{sugihara2012detecting} in their work.

EDM has been successfully applied to diverse research fields~\cite{chang2017empirical}. 
In neuroscience, CCM was applied to identify the effective connectivity between brain areas 
from magnetoencephalography (MEG) data~\cite{natsukawa2017visual}.
In ecology, Grziwotz \emph{et al.} found the causal relationships between the environment and 
mosquito abundance by using CCM~\cite{grziwotz2018empirical}. Environmental factors, such as
temperature, precipitation, dew point, air pressure, and mean tide level were identified to causally affect mosquito abundance.
Ma \emph{et al.} applied simplex projection to forecast wind generation~\cite{ma2017ultra}.
In~\cite{anderson2008fishing}, an EDM algorithm called S-Map~\cite{sugihara1994nonlinear} was used to find the relationship 
between harvested and unharvested fish in terms of size, age, and others.
Luo \emph{et al.} applied CCM to estimate the causal relationships of user behavior in an online social network~\cite{luo2014causal}.
These use cases demonstrate the wide applicability of EDM to analyze nonlinear dynamical systems.

\subsection{cppEDM} \label{cppEDM}

cppEDM~\cite{cppedm} is the latest implementation of the EDM framework.
cppEDM is a general purpose C++ library used as a backend by rEDM~\cite{redm} and pyEDM~\cite{pyedm}, 
which are EDM implementations for the R and Python language, respectively. 

We have identified two major issues in cppEDM that hinder large-scale analysis 
on HPC systems: redundant computation and lack of GPU support.
Since cppEDM is a general purpose library, it provides a one-to-one cross mapping function to identify 
the causality between a selected combinations of time series variables.
The all-to-all cross mapping function is implemented by reusing the one-to-one cross mapping function.
This results in redundant computation. Additionally, cppEDM is a reference implementation of EDM; therefore, 
it is not optimized for a specific hardware architecture such as GPUs.
Furthermore, cppEDM suffers from significant load imbalance among workers because it performs static decomposition of the problem. In fact, a performance evaluation in a previous work showed that the runtime of workers varied greatly from 5 hours to 8.5 hours~\cite{park2019massively}.

\section{mpEDM}\label{proposal}


In this section, we first outline the original causal inference algorithm in cppEDM.
Then, we describe the algorithmic improvement and the design of the inter-node and intra-node
parallelization in mpEDM.

\subsection{Original Algorithm} \label{original_algorithm}

\begin{algorithm}[t]
    \SetAlgoLined
    \DontPrintSemicolon
    \KwIn{Dataset $ts$ ($N$ time series of length $L$), maximum embedding dimension $E_{max}$}
    \KwOut{$N \times N$ causal map \textrho}
    \tcp{Phase 1: Simplex projection}
    \For{$i \leftarrow 1$ \KwTo $N$}{
        \For{$E \leftarrow 1$ \KwTo $E_{max}$}{
            $library \leftarrow$ First half of $ts[i]$\;
            $target \leftarrow$ Second half of $ts[i]$\;
            $indices, distances \leftarrow$ kNN$(library, target, E)$\;
            $distances\leftarrow$normalize$(distances)$\;
            $prediction \leftarrow$ lookup$(indices, distances, library, E)$\;
            \textrho$[E] \leftarrow$ corrcoef$(target, prediction)$\;
        }
        $optE[i] \leftarrow$ $\argmax\limits_{E}~$\textrho$[E]$\;
    }
    \tcp{Phase 2: CCM}
    \For{$i \leftarrow 1$ \KwTo $N$}{
        \For{$j \leftarrow 1$ \KwTo $N$}{
            $indices, distances \leftarrow$ kNN$(ts[i], ts[i], optE[j])$\;
            $distances\leftarrow$normalize$(distances)$\;
            $prediction \leftarrow$ lookup$(indices, distances, ts[j], optE[j])$\;
            \textrho$[i, j] \leftarrow$ corrcoef$(ts[j], prediction)$\;
        }
    }
    \caption{Causal Inference in cppEDM}\label{pseudo:cppedm}
\end{algorithm}

Algorithm~\ref{pseudo:cppedm} outlines the causal inference algorithm in cppEDM.
The input to the algorithm is an $L \times N$ array $ts$, where $L$ is the number of time steps within a time series and 
$N$ is the number of time series. In addition to 
the input dataset, maximum embedding dimension $E_{max}$ and time lag \texttau~need to be supplied.
The output is an $N \times N$ casual map \textrho.
The algorithm consists of two phases: (1) simplex projection and (2) CCM. 
Simplex projection finds the optimal embedding dimension for each time series.
CCM estimates the causal relationship between two time series
using the optimal embedded dimension obtained in the first phase.
Note that in the original definition of CCM, predictions are made multiple
times using randomly subsampled library sets of different sizes and it is
tested whether increasing the library set size improves the prediction
accuracy. In this research, we excluded this step since the convergence
test passes in most cases if the prediction using the full library set
achieves high accuracy.

In the first phase, simplex projection (line 1--11) takes a time series in the
dataset and splits into $library$, the first half, and $target$, the
second half (line 3--4). Next, both library and target are embedded into
$E$-dimensional space using time delayed embeddings. A $k$-nearest neighbors
(kNN) search is performed in the state space to find the $E+1$ nearest target
points from each library point (line 5). The search results are stored in
two lookup tables $indices$ and $distances$, both of which are two-dimensional
arrays of shape $L \times (E+1)$. Element $(i, j)$ in the indices array is the
index of the $j$-th nearest target point from library point $i$, whereas
element $(i, j)$ in the $distances$ array is the Euclidean distance between
the library point $i$ and its $j$-th nearest target point. The $distances$
array is then converted to exponential scale and each row is normalized (line
6). A one step ahead prediction of a target point is made by (1) obtaining the
indices of its $E+1$ library neighbors from $indices$, (2) obtaining the one
step ahead values of those library points from $library$ and (3) computing a weighted average
of the future library points using $distances$ (line 7). Finally, Pearson's
correlation coefficient is computed to evaluate the predictive skill of the
simplex projection using the prediction results and real observed withheld
values (line 8). This is repeated for every $E$ ranging from 1 to $E_{max}$ 
($\leq$20 in practice). The $E$ value that achieves the highest accuracy is
determined to be the optimal embedding dimension for the time series and stored
in $optE$ (line 10).

In the second phase, CCM (line 12--19) works similar to simplex projection but predicts between two different time series. A given $library$ time series is used to
cross predict another $target$ time series in the dataset to evaluate whether
the latter is the cause of the former. It computes and normalizes the kNN
tables from the library time series (line 14--15) and uses the tables to
predict the target time series (line 16). Note that simplex projection predicts within the same time series while CCM predicts across two different time series. Therefore, the kNN tables computed in the simplex projection phase cannot be reused in the CCM phase. The correlation between the predicted
values and the actual values represents strength of causality (line 17). In
this manner, causal inference is performed for all combinations of time series
in the dataset.

We have profiled cppEDM and found out that over 97\% of the total runtime is
spent in the kNN search. In addition, we have discovered that the time delayed
embedding in cppEDM replicates the time series $E+1$ times and causes significant memory
overhead.

\subsection{Improved Algorithm} \label{improved_algorithm}


The key observation behind our algorithmic improvement is that the kNN lookup
table for CCM is constructed from the library time series only, and
the target time series is not used. This suggests that once the kNN lookup table
is computed for a particular library time series, we can reuse the precomputed table to
make predictions for every target time series. This improvement is
trivial if $N$ is in the same order as $E_{max}$, which was the case in
previous use cases of EDM. However, in our use case $N$ is equal to the
number of active neurons in a zebrafish brain, which is roughly $10^5$.
Therefore, the potential speedup becomes significantly large.

Algorithm~\ref{pseudo:mpedm} shows the pseudocode of the improved causal inference algorithm in mpEDM. The simplex projection algorithm is unchanged from cppEDM but its kNN and lookup functions are parallelized and optimized.
The CCM algorithm in mpEDM is improved in the following manner. For each library
time series, we first compute the kNN lookup tables for every embedding
dimension ranging from 1 to $E_{max}$ (line 4--7). Then, we iterate through all
$target$ time series and use the precomputed lookup table for the optimal
embedding dimension of the $target$ time series to predict the $target$ time series
(line 9--10). Finally, we compute the correlation between the prediction and
the actual $target$ to estimate the causality (line 11).

\begin{algorithm}[t]
    \SetAlgoLined
    \DontPrintSemicolon
    \KwIn{Dataset $ts$ ($N$ time series of length $L$), maximum embedding dimension $E_{max}$}
    \KwOut{$N \times N$ causal map \textrho}
    \tcp{Phase 1: Simplex projection}
    \For{$i \leftarrow 1$ \KwTo $N$}{
        \tcp{Same as cppEDM (Algorithm~\ref{pseudo:cppedm})}
    }
    \BlankLine
    \tcp{Phase 2: CCM}
    \For{$i \leftarrow 1$ \KwTo $N$}{
        \For{$E \leftarrow 1$ \KwTo $E_{max}$}{
            $indices[E], distances[E] \leftarrow$ kNN$(ts[i], ts[i], E)$\;
            $distances\leftarrow$normalize$(distances)$\;
        }
        \For{$j \leftarrow 1$ \KwTo $N$}{
            $E_j \leftarrow optE[j]$\;
            $prediction \leftarrow$ lookup$(indices[E_j], distances[E_j], ts[j], E_j)$\;
            \textrho$[i, j] \leftarrow$ corrcoef$(ts[j], prediction)$\;
        }
    }
    \caption{Causal Inference in mpEDM}
    \label{pseudo:mpedm}
\end{algorithm}

Algorithms~\ref{pseudo:knn_cpu} outlines the kNN function for CPU. We first calculate the
all-to-all distances between every library and target point in the state
space. Note that we do not explicitly create the time series embeddings on
memory but we compute them on-the-fly to reduce memory footprint and increase
cache hit. In addition, both $indices$ and $distances$ are stored in row-major
format to match the access pattern. Then, each row in the $distances$ and
$indices$ arrays is partially sorted in descending order using the distances
as sort keys. We use heap sort to implement partial sort. After the sorting,
both arrays are trimmed from $L \times L$ to $L \times (E+1)$ and returned.
Algorithm~\ref{pseudo:knn_gpu} shows the kNN function for GPU. In the GPU version, we create time
series embeddings on the host and transfer them to the device. The kNN search
is executed on the GPU and the resulting kNN tables are returned to the host.

Algorithm~\ref{pseudo:lookup} outlines the lookup function. It uses the kNN lookup tables $indices$ and $distances$ of the $library$ time series. 
For each target point, the indices of its $E+1$ neighbors are retrieved from the $indices$ table. 
Then, those neighbors are accumulated using the weights stored in the $distances$ table. 
Finally, the function returns the predicted $target$ time series.

The average time complexity of each algorithm is analyzed as follows.
The time complexity of the kNN function in Algorithm~\ref{pseudo:knn_cpu} and~\ref{pseudo:knn_gpu} is $O(L^2E)$ 
because the all-to-all distance calculation is $O(L^2E)$ and the sorting is approximately $O(L^2\log E)$.
The time complexity of the lookup function in Algorithm~\ref{pseudo:lookup} is $O(LE)$. 
By combining these results, the time complexity of simplex projection in mpEDM is $O(NL^2E)$, which is the same as cppEDM. 
The time complexity of CCM in mpEDM, on the other hand, is $O(NL^2E^2+N^2LE)$.
In cppEDM, the time complexity of CCM is $O(N^2L^2E)$. As a result, the time complexity of the whole causal 
inference algorithm in mpEDM is $O(NL^2E^2+N^2LE)$.

\begin{algorithm}[t]
    \SetAlgoLined
    \DontPrintSemicolon
    \KwIn{$library$ and $target$ time series, embedding dimension $E$, time lag \texttau}
    \KwOut{Arrays $diatances$ and $indices$ for lookup}
    \tcp{All-to-all distance calculation}
    \For{$i \leftarrow 1$ \KwTo $L$}{
        \For{$k \leftarrow 1$ \KwTo $E$}{
            $distances[i, :] \leftarrow 0$\;
            \For{$j \leftarrow 1$ \KwTo $L$}{
                $indices[i, j] \leftarrow j$\;
                $distances[i, j] \leftarrow distances[i, j] + (target[k $\texttau$ + i] - library[k $\texttau$ + j])^2$\;
            }
        }
    }
    \tcp{Sorting}
    $top\_k \leftarrow$ E+1\;
    \For{$i \leftarrow 1$ \KwTo $L$}{
        $indices[i, :] \leftarrow $ partialSort$(indices, distances, top\_k)$\;
    }
    \caption{kNN for CPU}
    \label{pseudo:knn_cpu}
\end{algorithm}

\begin{algorithm}[t]
    \SetAlgoLined
    \DontPrintSemicolon
    \KwIn{$library$ and $target$ time series, embedding dimension $E$, time lag \texttau}
    \KwOut{Arrays $diatances$ and $indices$ for lookup}
    \tcp{Embedding}
    \For{$i \leftarrow 1$ \KwTo $E$}{
        \For{$j \leftarrow 1$ \KwTo $L$}{
            $libraryBlock[i, j] \leftarrow library[i $\texttau$ + j]$\;
            $targetBlock[i, j] \leftarrow target[i $\texttau$ + j]$\;
        }
    }
    \tcp{All-to-all distance calculation and sorting}
    $top\_k \leftarrow$ E+1\;
    Copy $libraryBlock$ and $targetBlock$ to device\;
    $indices, distances \leftarrow$ nearestNeighbour$(libraryBlock, targetBlock, top\_k)$\;
    Copy $indices$ and $distances$ to host\;
    \caption{kNN for GPU}
    \label{pseudo:knn_gpu}
\end{algorithm}

\begin{algorithm}[t]
    \SetAlgoLined
    \DontPrintSemicolon
    \KwIn{Array of $indices$ and $distances$, $target$ time series, embedding dimension $E$ of $target$}
    \KwOut{Prediction of the time series $prediction$}
    \For{$i \leftarrow 1$ \KwTo $L$}{
        $prediction[i] \leftarrow 0$\;
        \For{$j \leftarrow 1$ \KwTo $E+1$}{
            $idx \leftarrow indices[i, j]$\;
            $dist \leftarrow distances[i, j]$\;
            $prediction[i] \leftarrow prediction[i] + target[idx] \cdot dist$\;
        }
    }
    \caption{Lookup}
    \label{pseudo:lookup}
\end{algorithm}

\subsection{Inter-Node Parallelism}\label{inter-node-parallelism}


To distribute the work across multiple compute nodes, we naturally choose the
loops with the highest granularity. That is, the two outermost loops that
iterate over the time series (line 1--2 and 3--13 in Algorithm~\ref{pseudo:mpedm}). We
implement a simple master-worker framework based on MPI to distribute these
loops. To dynamically distribute work and mitigate load imbalance among workers,
we adopt \emph{self-scheduling} in our master-worker framework. In self-scheduling,
the master accounts and dispatches tasks to workers. Each worker performs
assigned tasks, and once it completes, the worker asks the master for a new
task.

The high-level organization of the inter-node parallelism is as follows.
First, the workers execute the simplex projection phase. The optimal
embedding dimension for each time series is reported back to the master. Once the
first phase is complete, the master broadcasts $optE$ to all workers.
Subsequently, the workers execute the all-to-all CCM phase. The
final results are written to the file system by each worker to alleviate the load on
the master.

Both the input dataset and the inferred causal map are stored as HDF5~\cite{folk1999hdf5} files
for easy integration with the pre/post processing workflow. The workers read
the input HDF5 file in parallel and keep the entire dataset on memory during
the execution. Every time a worker completes a cross map, the worker writes
an element of the causal map asynchronously to the output HDF5 file. This small
random write pattern, however, is known to be slow on parallel
file systems. In fact, we observed that write I/O becomes a significant
bottleneck of the application on GPFS. We therefore take advantage of BeeOND (BeeGFS
On Demand)~\cite{beeond}, the burst buffer deployed on ABCI. BeeOND combines local SSDs
installed on the compute nodes and provides an on-demand parallel file system
to a job. The workers write the results to BeeOND to minimize I/O overhead.

\subsection{Intra-Node Parallelism}\label{intra-node-parallelism}

We focus our efforts to parallelize and optimize the kNN kernel since it is
the primary bottleneck in cppEDM as discussed in section~\ref{original_algorithm}. We design and
implement kNN kernels for both CPU and GPU architecture to ensure that mpEDM
can efficiently run on a wide variety of computing platforms. 
In the kNN kernel for CPU shown in Algorithm~\ref{pseudo:knn_cpu}, the two loops that
iterate over the time steps within a time series are parallelized using OpenMP
(line 1--9 and 10--13 in Algorithm~\ref{pseudo:knn_cpu}). We also utilize OpenMP 4.0 SIMD
directives to vectorize the innermost loop explicitly. Note that the nested loops
are ordered such that the memory accesses in the innermost loop are
contiguous.

In the kNN kernel for GPU shown in Algorithm~\ref{pseudo:knn_gpu}, we take advantage
of ArrayFire\cite{malcolm2012arrayfire}, a highly optimized library for GPU-accelerated computing.
ArrayFire provides backends for CUDA, OpenCL and CPU, but in this paper we
only use the CUDA backend since ABCI is installed with Tesla V100 GPUs. The
kNN algorithm implemented in ArrayFire is essentially the same as our CPU
implementation. ArrayFire uses a block-wide parallel radix sort implementation
in the CUDA UnBound (CUB) template library. Since each ABCI compute node
is equipped with four GPUs, we also distribute the work across multiple GPUs.
To achieve this, the loop that iterates over $E$ (line 4--7 in Algorithm~\ref{pseudo:mpedm}) is
parallelized such that each GPU computes lookup tables for one or more $E$. We dynamically
schedule this loop to ensure load balancing across GPUs because the runtime of
the kNN kernel depends on $E$ as discussed in section~\ref{improved_algorithm}.

For the lookup kernel shown in Algorithm~\ref{pseudo:lookup}, we currently only have a
CPU version of this kernel. The time step loop is parallelized using OpenMP
(line 1--8 in Algorithm~\ref{pseudo:lookup}). This kernel is heavily memory bandwidth bound since it
requires random memory access.

\section{Evaluation} \label{evaluation}

The computational performance of mpEDM was evaluated on
ABCI. Furthermore, we present the scientific outcomes 
obtained using mpEDM.

\subsection{Evaluation Environment}

ABCI~\cite{abci} is the world's first large-scale Open
AI Computing Infrastructure, which is constructed and operated by the National
Institute of Advanced Industrial Science and Technology (AIST). According to
the latest TOP500 list published in November 2019~\cite{top500nov19}, ABCI is
the most powerful supercomputer in Japan and the 8th in the world. ABCI has
1,088 compute nodes, each equipped with two 20-core Intel Xeon Gold 6148 CPUs,
four NVIDIA Tesla V100 SXM2 (16GB) GPUs, 384GB of RAM  and 1.6TB of local NVMe SSD. 
The parallel file system is based on GPFS with a total capacity of 22PB.

\subsection{Performance Evaluation} \label{performance_evaluation}


We compared mpEDM with cppEDM from the following three aspects: total runtime, 
parallel scalability and impact of dataset size on the runtime.
We used three real-world datasets recorded from larval zebrafish under different conditions.
Table~\ref{table:dataset} shows the list of datasets used in the evaluation.


\begin{table}[htbp] 
    \centering
    \caption{Datasets used in the evaluation}
    \begin{tabular}{@{}lrrl@{}}
        \toprule
        \multicolumn{1}{c}{Dataset} & \multicolumn{1}{c}{\# of Time Steps} & \multicolumn{1}{c}{\# of Time Series} & Size   \\ \midrule
        Fish1\_Normo                & 1,450                                 & 53,053                                & 0.7 GB \\
        Subject6                    & 3,780                                 & 92,538                                & 3.0 GB \\
        Subject11                   & 8,528                                 & 101,729                               & 9.5 GB \\ \bottomrule
    \end{tabular}
    \label{table:dataset}
\end{table}


\subsubsection{Total Runtime}

mpEDM shows significantly higher performance compared to cppEDM.
Table~\ref{table:mpedm_cppedm_result} shows the performance comparison between
cppEDM and mpEDM. cppEDM took 8.5 hours to analyze the Fish1\_Normo dataset
using 512 ABCI nodes~\cite{park2019massively}, whereas mpEDM took only 20 seconds to analyze the same
dataset using 512 ABCI nodes with GPU architecture. The result shows that mpEDM is 1,530$\times$ faster than cppEDM.
Moreover, mpEDM finished the causal inference of two larger datasets: Subject6 in 101 seconds and Subject11~\cite{chen2018brain} 
in 199 seconds.

\begin{table}[htbp]
    \centering
    \caption{Performance comparison between cppEDM and mpEDM}
    \begin{tabular}{@{}lrrr@{}}
        \toprule
        \multicolumn{1}{c}{}        & \multicolumn{1}{c}{cppEDM}    & \multicolumn{2}{c}{mpEDM}                                     \\ \cmidrule(l){2-2} \cmidrule(l){3-4} 
        \multicolumn{1}{c}{Dataset} & \multicolumn{1}{c}{512 Nodes} & \multicolumn{1}{c}{1 Node}    & \multicolumn{1}{c}{512 Nodes} \\ \midrule
        Fish1\_Normo                & 8.5h                          & 1,973s                        & 20s                           \\
        Subject6                    & N/A                           & 13,953s                       & 101s                          \\
        Subject11                   & N/A                           & 39,572s                       & 199s                          \\ \bottomrule
    \end{tabular}
    \label{table:mpedm_cppedm_result}
\end{table}


\subsubsection{Parallel Scalability}


We measured the parallel scalability of mpEDM by varying the number of workers and measuring the runtime of mpEDM with and without GPU. 
We used the largest Subject11 dataset in this evaluation. 

Figure~\ref{fig:scale_result} shows the strong scaling performance of mpEDM.
In the \textit{Single Node} setup, mpEDM is executed on a single node without MPI. In the \textit{$X$ Workers} setup, mpEDM is executed
with MPI using the specified number of workers.
We measured up to 511 workers since ABCI allows a maximum of 512 nodes per job (except for jobs running under the ABCI grand challenge program, which can use the full 1,088 nodes).
The result shows that the GPU version runs as twice as fast as the CPU version
in every case. We noticed that the CPU version ran in the single worker setup 10\% slower than the single node setup. 
We believe this slowdown is caused from the interference between the background tasks performed by the BeeOND daemon and the computation in mpEDM.
This does not happen with the GPU version because the average CPU utilization is lower than the CPU version.

Figure~\ref{fig:scale_speedup} shows the relative speedup of the multi-node setup in relation to the single node setup.
It reveals that the speedup is nearly linear with both GPU and CPU. However, the speedup of the GPU version drops when the number of nodes is 64 or more.

\begin{figure}
    \centering
    \centerline{\includegraphics[scale=0.5]{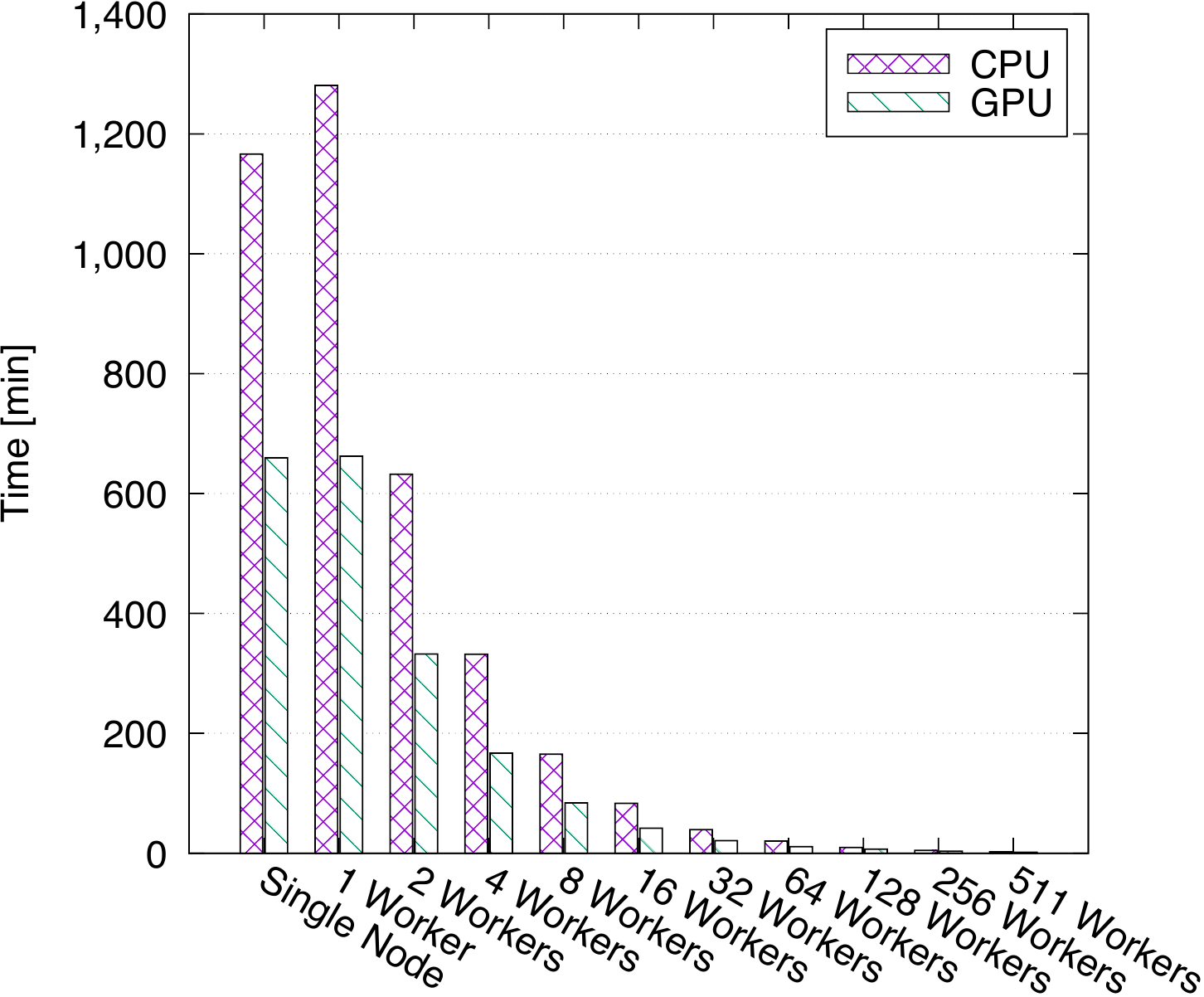}}
    \caption{Strong scaling performance (absolute runtime)}
    \label{fig:scale_result}
\end{figure}

\begin{figure}
    \centering
    \centerline{\includegraphics[scale=0.5]{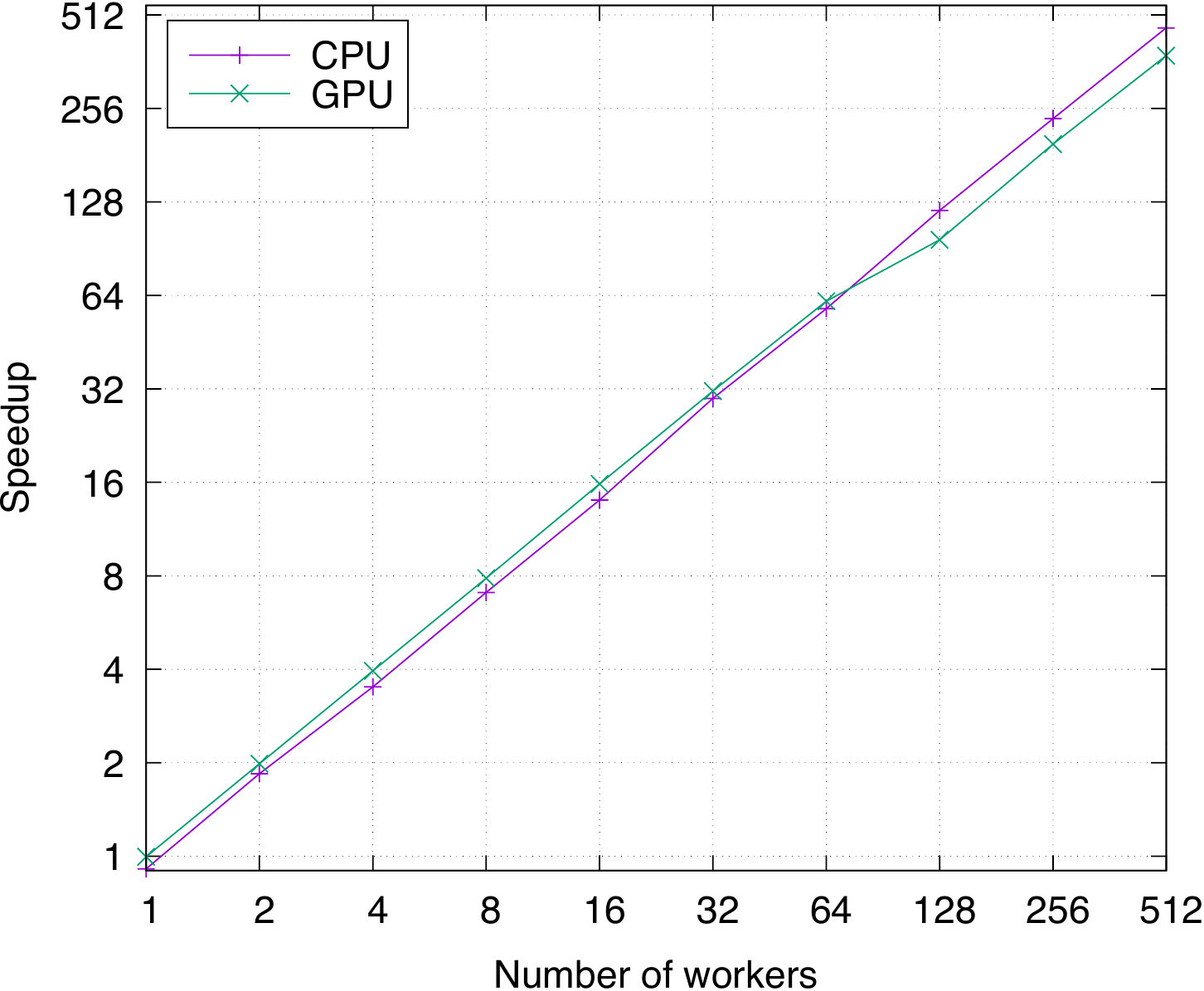}}
    \caption{Strong scaling performance (relative speedup)}
    \label{fig:scale_speedup}
\end{figure}

\begin{figure}[t]
    \centering
    \centerline{\includegraphics[scale=0.5]{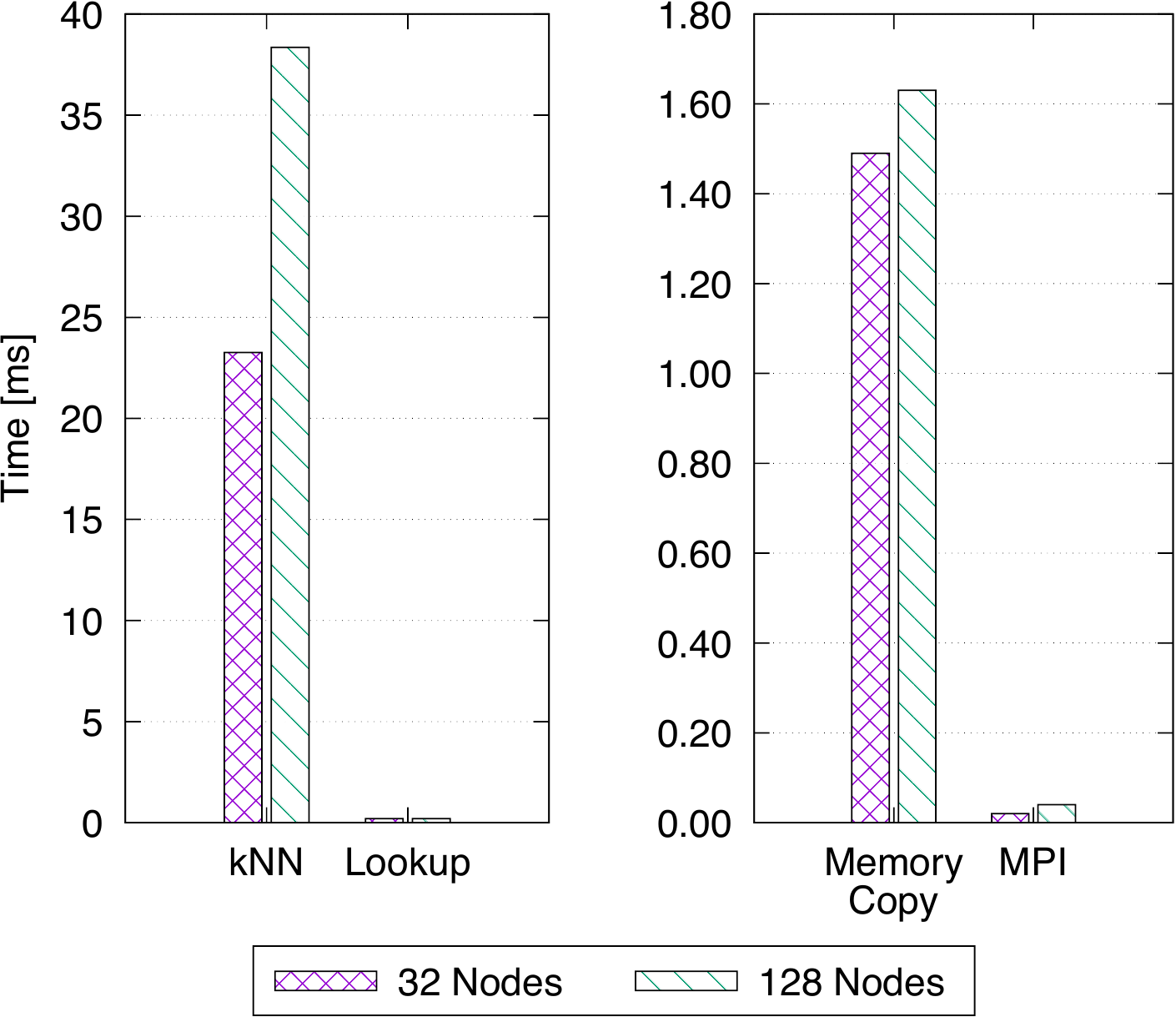}}
    \caption{Breakdown of simplex projection (average runtime per time series)}
    \label{fig:simplex_breakdown}
\end{figure}

\begin{figure}[t]
    \centering
    \centerline{\includegraphics[scale=0.5]{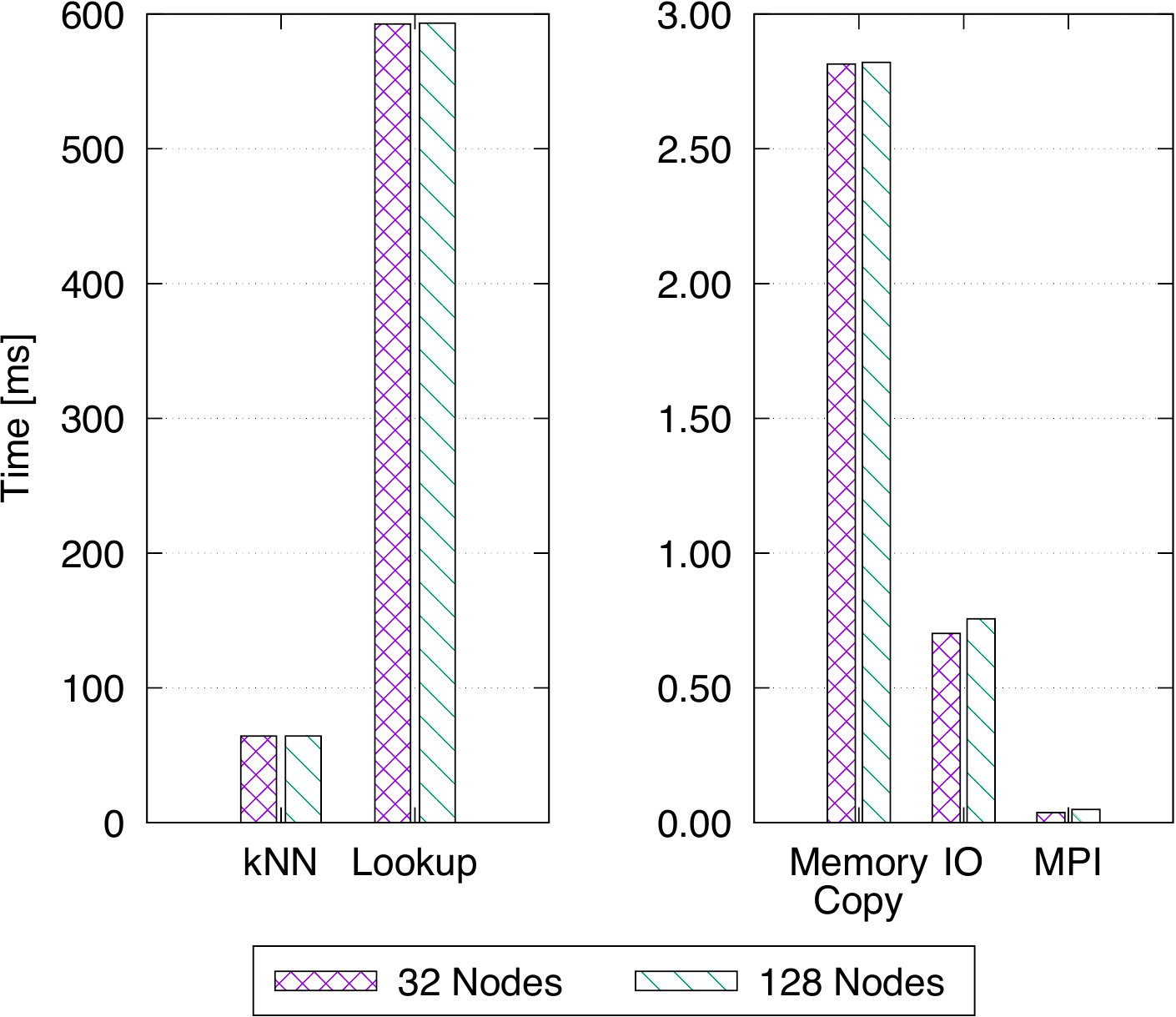}}
    \caption{Breakdown of cross mapping (average runtime per time series)}
    \label{fig:crossmap_breakdown}
\end{figure}



\begin{figure}
    \centering
    \centerline{\includegraphics[scale=0.5]{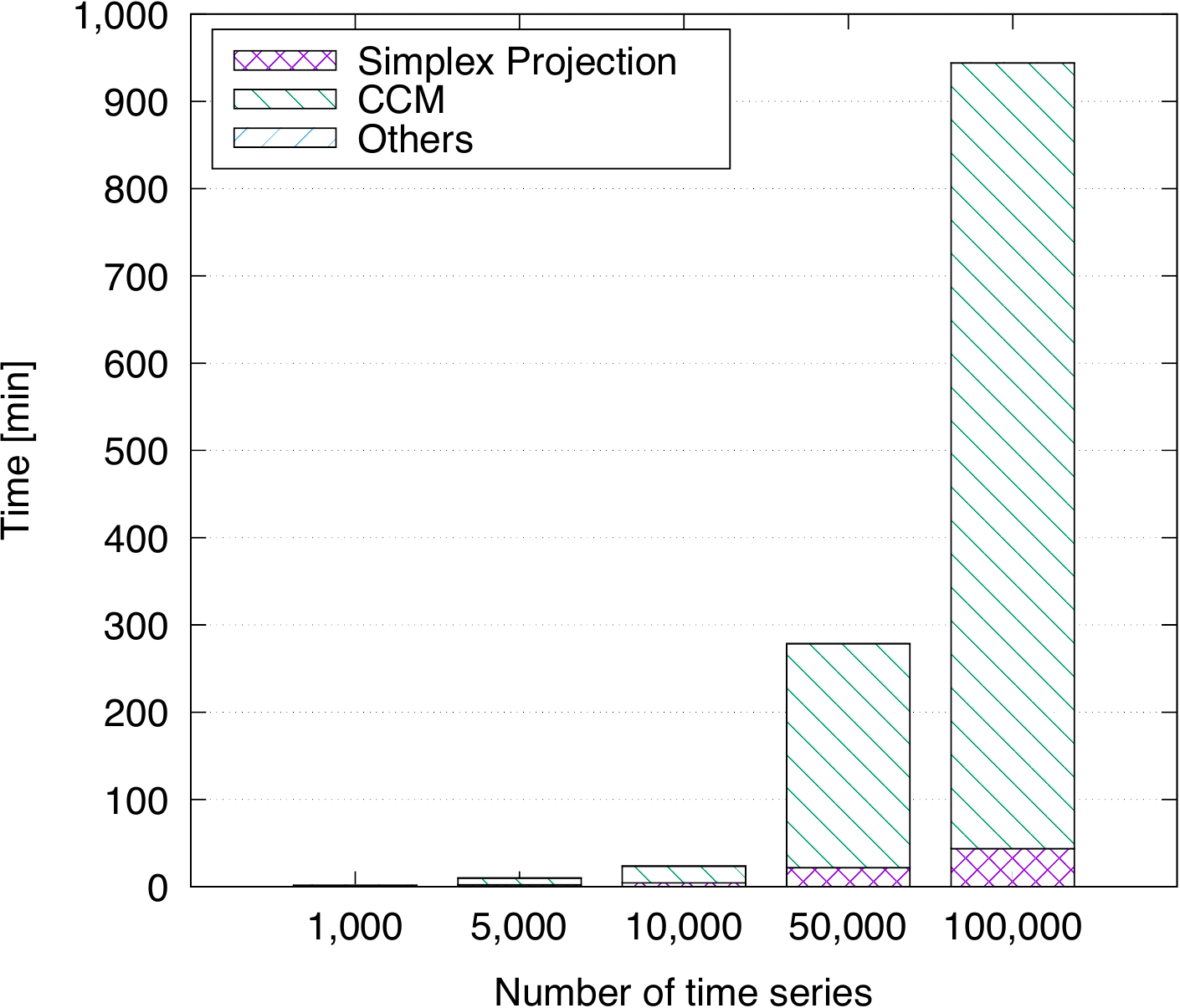}}
    \caption{Runtime with varying number of time series (10,000 time steps)}
    \label{fig:dataset_scale_timeseries}
\end{figure}

\begin{figure}
    \centering
    \centerline{\includegraphics[scale=0.5]{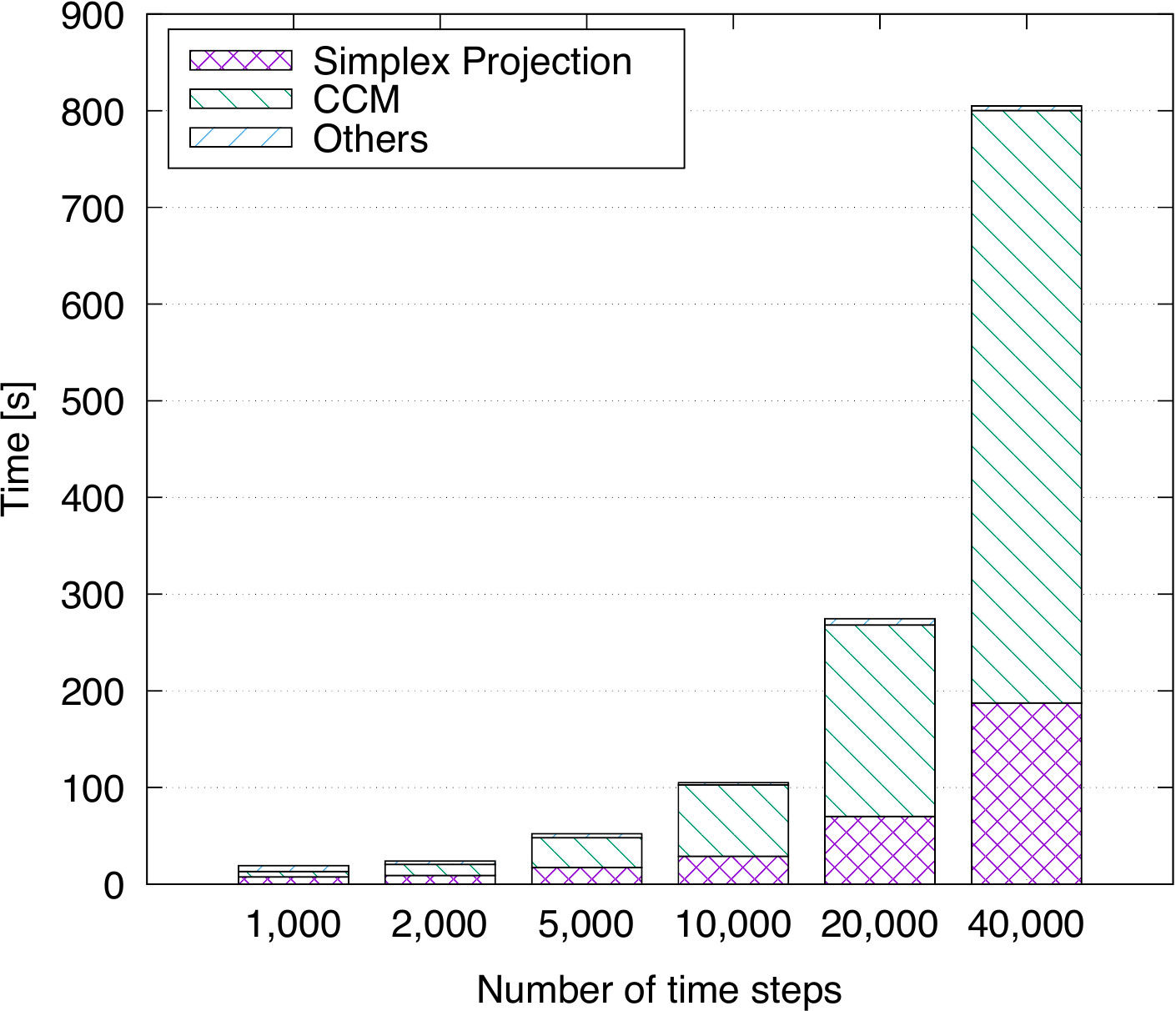}}
    \caption{Runtime with varying number of time steps (1,000 time series)}
    \label{fig:dataset_scale_timepoints_gpu}
\end{figure}

We measured the breakdown of each phase to investigate the cause behind the scalability decline. We compared 32 workers and 128 workers since the GPU version declines beyond 64 nodes. Figures~\ref{fig:simplex_breakdown} and~\ref{fig:crossmap_breakdown} show the breakdown of average runtime for processing a single time series in simplex projection and CCM. The two figures clearly indicate that memory copy, MPI communication and I/O are not bottlenecks and do not
significantly increase with the number of workers. However, the kNN function becomes slower when the number of workers increases. We found out that the kNN search for the first time series processed on a worker is significantly slower (ranging from 3.3 seconds to 16.4 seconds) than the subsequent ones. We believe this is caused by the initialization process of the GPUs.



\begin{figure}
    \centering
    \centerline{\includegraphics[scale=0.5]{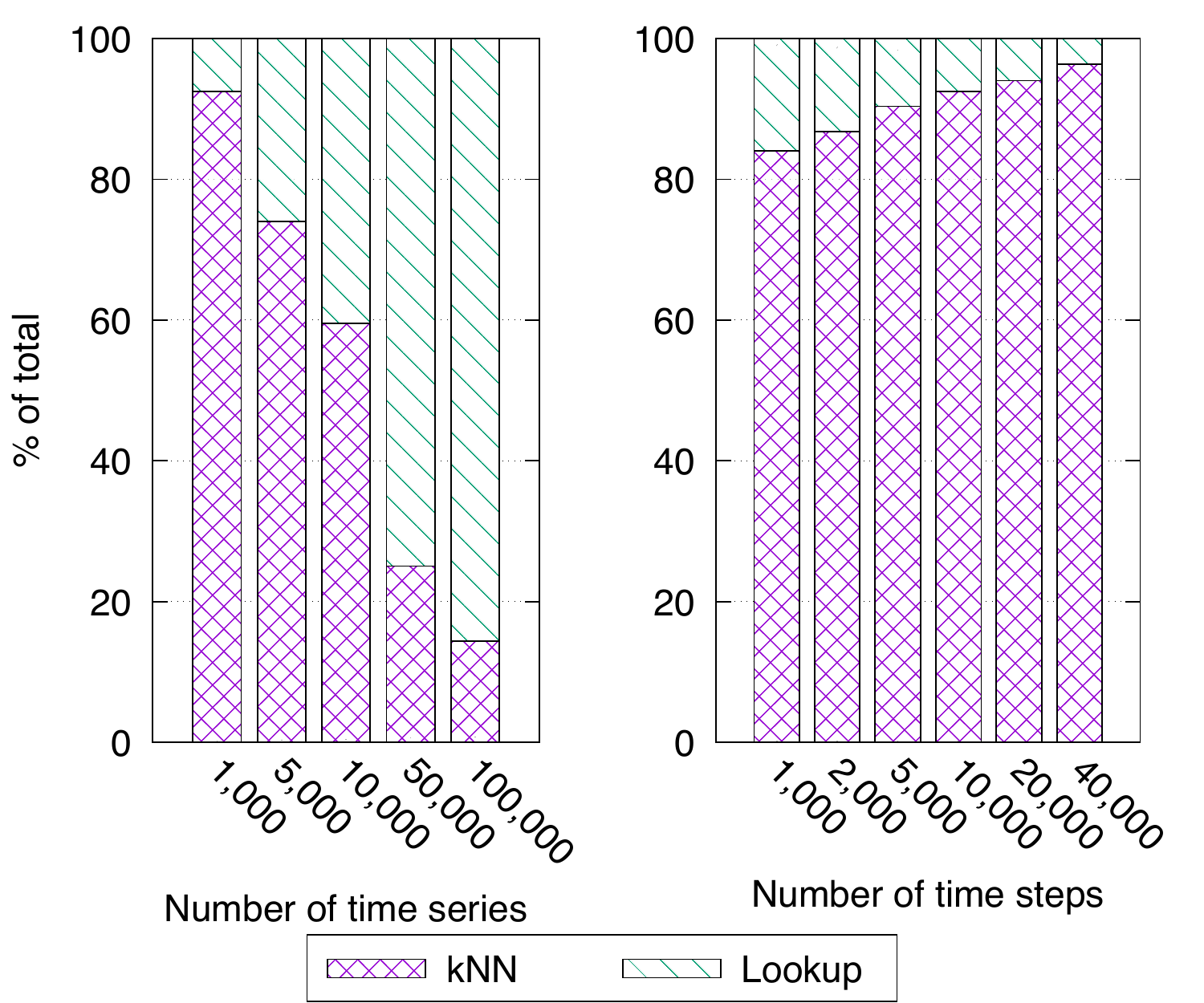}}
    \begin{subfigure}{.48\linewidth}
        \medskip
        \caption{Varying number of time series (10,000 time steps)}
        \label{fig:dataset_scale_cm_timeseries}
    \end{subfigure}
    \begin{subfigure}{.48\linewidth}
        \medskip
        \caption{Varying number of time steps (1,000 time series)}
        \label{fig:dataset_scale_cm_timepoints}
    \end{subfigure}
    \caption{Breakdown of CCM}
    \label{fig:dataset_scale_cm}
\end{figure}  

\begin{figure}
    \centering
    \centerline{\includegraphics[scale=0.5]{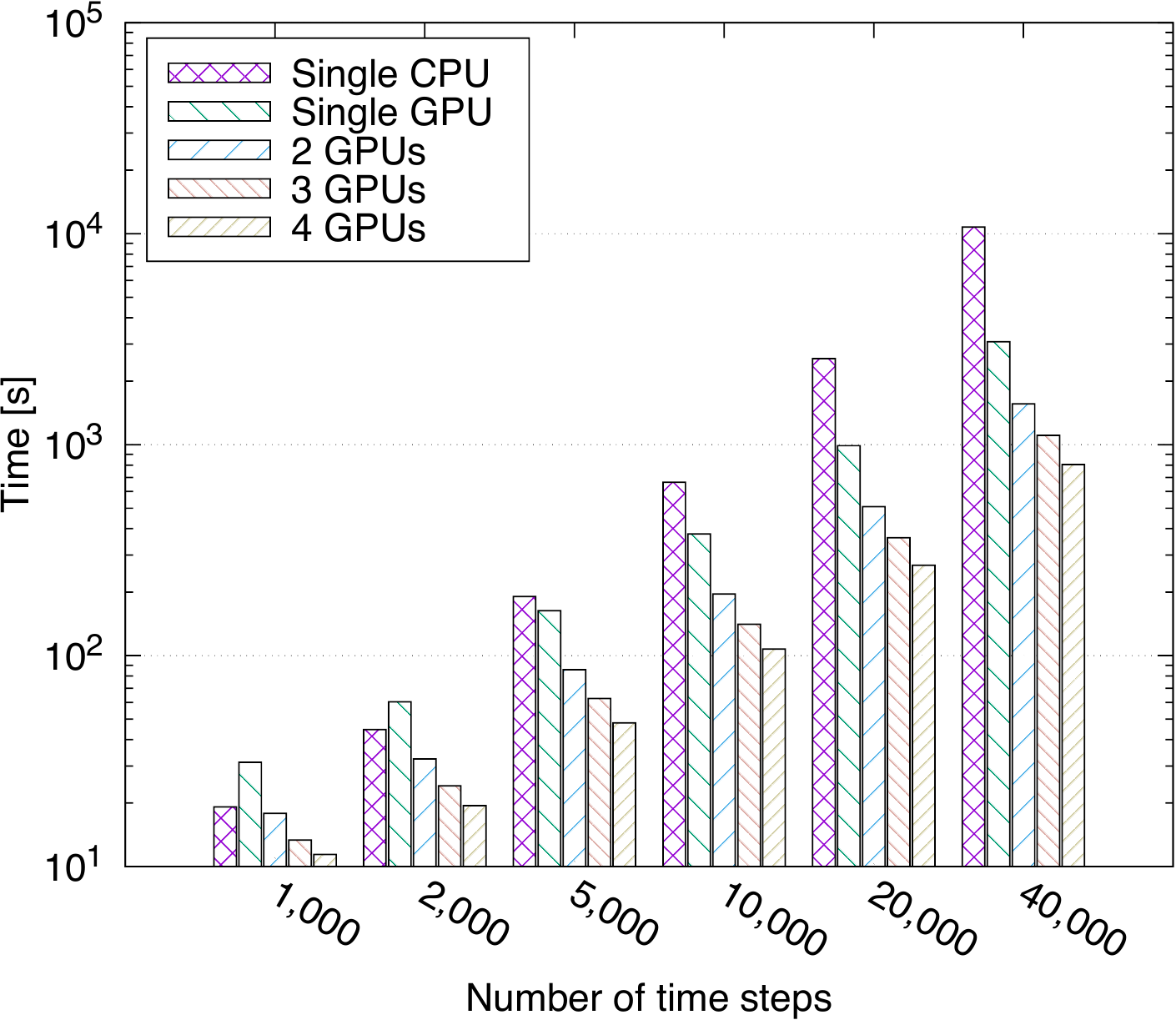}}
    \caption{GPU speedup with varying number of time steps (1,000 time series)}
    \label{fig:dataset_scale_timepoints_cpu_gpu_speedup}
\end{figure}

\begin{figure*}
    \centering
    \centerline{\includegraphics[width=1\textwidth]{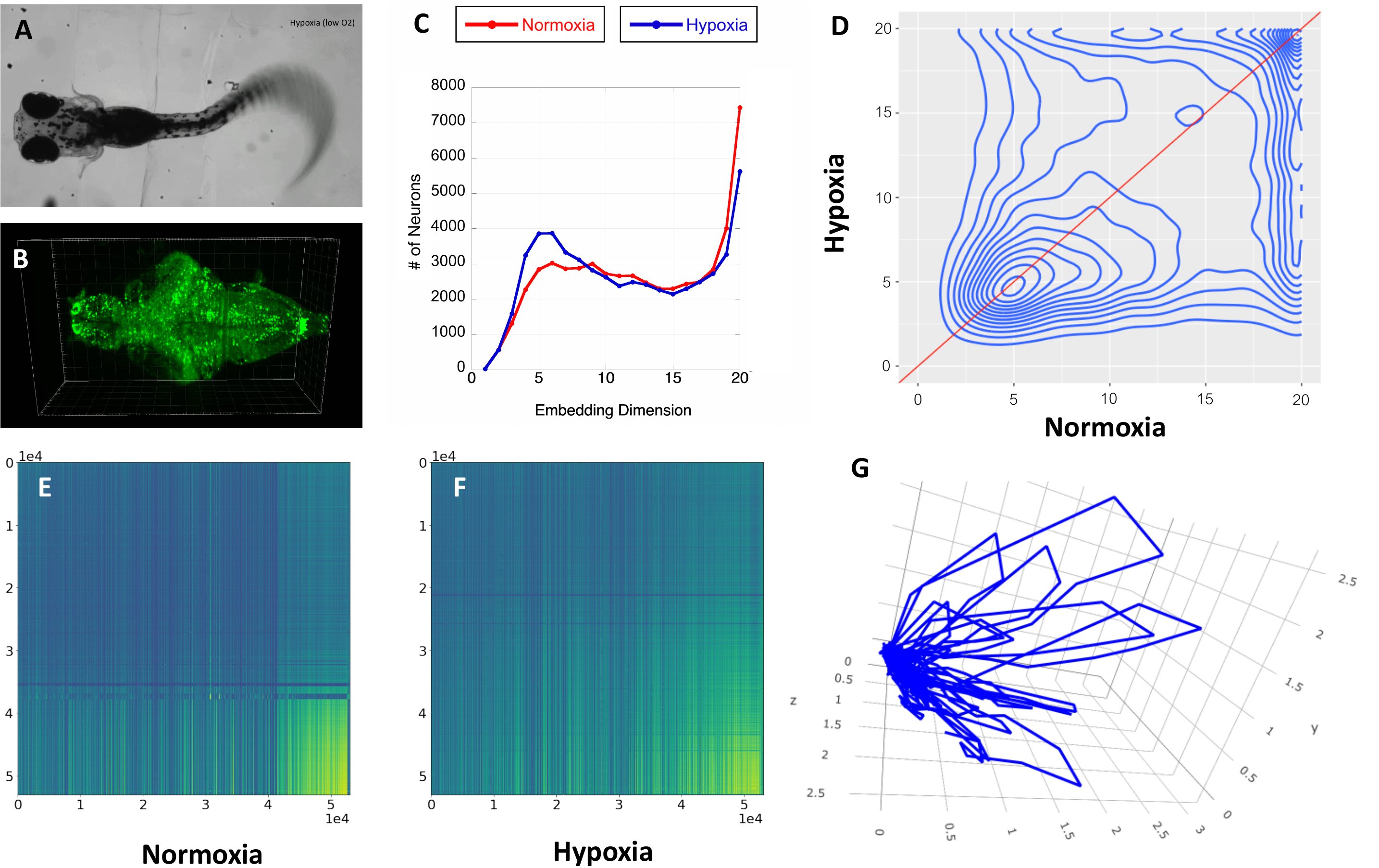}}
    \caption{Scientific results. 
    (A) Zebrafish larvae were imaged to study their response to low oxygen 
    (B) The day larvae were imaged using a SPIM lightsheet microscope and whole brain calcium activity was recorded at single cell resolution 
    (C) Our calculation of the dimensionality of the neuronal populations show a decrease under low oxygen (hypoxia) as seen in the distribution.
    (D) Measured transitions between normal oxygen concentrations (normoxia) to hypoxia show a bias below to the right of the diagonal line showing that dimensionality decreases as oxygen decreases. (E, F) Whole brain CCM all vs all causal inference matrix of an all vs all neurons. Results show a more homogeneous map in hypoxia 
    (F) than normoxia (E) indicating a simplification of behavior consistent with the above dimensionality drop. 
    (G) An identified signal integration manifold capable of predicting turns of the fish at least 0.5 seconds (a single time step) ahead of time. Whenever the neural activity trajectory enters one of the loops of the manifold, the fish will turn.
    }
    \label{fig:scientific_result}
\end{figure*}

To verify this, we created a simple program that initializes the GPUs and allocates some GPU memory on a single node. We submitted a job that run this program 100 times and measured the initialization time. The result revealed that the initialization time follows a long-tailed distribution: the median was 4.6 seconds while the maximum was 22.9 seconds. This suggests that a few stragglers impact the total runtime and degrade the scalability as the number of workers increases.


\subsubsection{Impact of Dataset Size}

We evaluated how the size of the dataset impacts the runtime of mpEDM using
dummy datasets with different sizes. Furthermore, we measured the time spent
in each function. We also measured the speedup of the GPU version over the CPU version with varying number of time steps.

Figures~\ref{fig:dataset_scale_timeseries} and~\ref{fig:dataset_scale_timepoints_gpu} show
the runtime of mpEDM when increasing the number of time series and time steps, respectively.
We confirmed that the increase of runtime is not bigger than the increase predicted from the time complexity.
We also confirmed that CCM consumed the majority of the total runtime and other tasks including I/O and MPI communication are ignorable.

Figures~\ref{fig:dataset_scale_cm_timeseries} and~\ref{fig:dataset_scale_cm_timepoints} show the
runtime breakdown of each function in CCM when increasing the number of time series and time steps, respectively. 
Figure~\ref{fig:dataset_scale_cm_timeseries} shows that the runtime of the lookup
function becomes dominant when increasing the number of time series.
On the other hand, Fig.~\ref{fig:dataset_scale_cm_timepoints} shows that the runtime of the kNN
function becomes dominant when increasing the number of time steps.
These trends can be explained from the time complexity analysis of each algorithm described 
in section~\ref{improved_algorithm}. 

Figure~\ref{fig:dataset_scale_timepoints_cpu_gpu_speedup} shows the speedup of the GPU version 
over the CPU version when varying number of time steps. We compared the performance between a single CPU socket and one or more GPUs to evaluate the GPU speedup.
Evidently, the GPU speedup increases with the number of time steps.
Single GPU is slower than the CPU if the number of time steps is 2,000 or less. This is because of the overhead inherent to offloading computation to the GPU. However, single GPU consistently surpasses the CPU if the number of time steps if 5,000 or more. If the number of time steps is 40,000, the speedup of a single GPU is 3.5 times compared to CPU. When four GPUs are used, the speed up is 13.4 times.

\subsection{Scientific Outcomes}

Figure~\ref{fig:scientific_result} shows the scientific outcomes obtained using mpEDM. Our results showed that we could determine the causal connectivity across the entire brain across two behaviors. This shows that depending on task, the network of relationships between individual neurons change and become more connected, homogeneous and simplified with a goal directed task. In the resulting network connectivity increased and became simpler. Furthermore, we were able identify individual neurons that integrate signals from multiple other neurons that contain decision making information. These neurons allow the prediction of fish turn behaviors while swimming and generate low dimensional manifold models based on data geometry that are able to predict the fish’s behavior at least 0.5 seconds (a single time step) ahead. A three dimensional projection of one of these manifolds is shown in Fig.~\ref{fig:scientific_result} (G), where entering the loop predicts turn behavior. Based on the combined activity of two neurons and information on prior states we are able to predict when the fish will turn. Beyond this, this is the first map of causal connectivity of any vertebrate animal at single neuron resolution.

\section{Conclusion \& Future Work}\label{summary}

EDM is a nonlinear time series analysis framework
proven its applicability in various fields. However, EDM has only been applied to
small datasets due to its computational cost. In this paper, we designed and implemented mpEDM, a parallel distributed
implementation of EDM optimized for execution on modern GPU-centric
supercomputers. mpEDM improves the EDM algorithm to reduce redundant computation and optimizes the 
implementation to fully utilize hardware resources such as GPUs and SIMD units. 
mpEDM took only 20 seconds to finish the causal inference of a dataset containing the activity of 53,053 
zebrafish neurons on 512 ABCI nodes. This is 1,530$\times$ faster than cppEDM, the current standard implementation of EDM.
Moreover, mpEDM could analyze a 13$\times$ larger dataset in 199 seconds.
This is the largest EDM causal inference achieved to date.

We will continue to optimize the performance of mpEDM.
As discussed in section~\ref{performance_evaluation}, we need to improve the performance of the 
lookup as it becomes the primary bottleneck when we scale up the number of time series further.
We will also explore other efficient implementations of nearest neighbor search on GPUs.
Currently, mpEDM uses the exact kNN search implementation provided by ArrayFire.
There exist many studies on efficient Approximate Nearest Neighbor (ANN) search~\cite{chen2019robustiq,pan2011fast}.
However, it is unclear how ANN affects the accuracy of EDM predictions.
Another well-known approach is to use spatial indices such as KD-trees and Ball-trees to
accelerate kNN search~\cite{garcia2008fast, abbasifard2014survey}.

Additionally, EDM algorithms other than simplex projection and CCM will be implemented in
mpEDM to expand mpEDM to a standard implementation of EDM on
HPC systems. We will make this EDM library widely available to the community with a hope 
to assist scientists in need to analyze large-scale time series datasets of nonlinear dynamical systems.

\section*{Acknowledgment}

This work was supported by JSPS KAKENHI Grant Number JP20K19808 (KT) and an Innovation grant by the Kavli Institute for Brain and Mind (GMP). Computational resources of the AI Bridging Cloud Infrastructure (ABCI) were provided by the National Institute of Advanced Industrial Science and Technology (AIST).

\bibliographystyle{ieeetr}
\bibliography{cuEDM_escience2020}


\end{document}